# High-precision methanol spectroscopy with a widely tunable SI-traceable frequency comb-based mid-IR QCL

R. Santagata[1,2,†], D.B.A. Tran[1,§,†], B. Argence[1,#], O. Lopez[1], S. K. Tokunaga[1], F. Wiotte[1], H. Mouhamad[1], A. Goncharov[1,††], M. Abgrall[2], Y. Le Coq[2], H. Alvarez-Martinez[2,3], R. Le Targat[2], W. K. Lee[2,4], D. Xu[2], P.-E. Pottie[2], B. Darquié[1,*], A. Amy-Klein[1]

[1]*Laboratoire de Physique des Lasers, Université Paris 13, Sorbonne Paris Cité, CNRS, Villetaneuse, France*
[2]*LNE-SYRTE, Observatoire de Paris, Université PSL, CNRS, Sorbonne Université, Paris, France*
[3]*Real Instituto y Observatorio de la Armada, San Fernando, Spain*
[4]*Korea Research Institute of Standards and Science, Daejeon 34113, South Korea*
*Corresponding author: benoit.darquie@univ-paris13.fr*

*Abstract—* There is an increasing demand for precise molecular spectroscopy, in particular in the mid-infrared fingerprint window that hosts a considerable number of vibrational signatures, whether it be for modeling our atmosphere, interpreting astrophysical spectra or testing fundamental physics. We present a high-resolution mid-infrared spectrometer traceable to primary frequency standards. It combines a widely tunable ultra-narrow Quantum Cascade Laser (QCL), an optical frequency comb and a compact multipass cell. The QCL frequency is stabilized onto a comb controlled with a remote near-infrared ultra-stable laser, transferred through a fiber link. The resulting QCL frequency stability is below $10^{-15}$ from 0.1 to 10s and its frequency uncertainty of $4\times10^{-14}$ is given by the remote frequency standards. Continuous tuning over ~400 MHz is reported. We use the apparatus to perform saturated absorption spectroscopy of methanol in the low-pressure multipass cell and demonstrate a statistical uncertainty at the kHz level on transition center frequencies, confirming its potential for driving the next generation technology required for precise spectroscopic measurements.

## 1. INTRODUCTION

Precise spectroscopic analysis of various molecular systems enables many exciting advances in physical chemistry and fundamental physics. For example, accurate spectroscopic models are required for improving our understanding of atmospheric physics [1,2] and astro-chemistry [3], but applications also include industrial monitoring, environmental analysis, medical diagnosis or precise tests of fundamental physics [4–12]. The mid-infrared (MIR) spectral window, the so-called molecular fingerprint region, is of particular interest as it hosts intense vibrational signatures of a considerable number of species.

This work is dedicated to the development of a quantum cascade laser (QCL) based spectrometer that provides a unique combination of precision and tunability of any MIR spectrometer to date. Available in the whole MIR region, continuous wave QCLs offer broad and continuous tuning over several hundred gigahertz at milliwatt to watt-level powers, but show substantial free-running frequency fluctuations. For the most precise frequency measurements considered here, frequency stabilization and traceability to a frequency standard are both required [13–20]. As demonstrated in [16], this can be achieved by phase-locking to the secondary frequency standard of this spectral region, a $CO_2$ laser stabilized on a molecular saturated absorption line [21,22]. Very few such standards are however only available around 10 µm, and their stability and accuracy are a few orders of magnitude worse than the state-of-the-art performances offered by the best near-infrared (NIR) ultra-stable lasers [23] and atomic references respectively. These are available in national metrological institutes (NMI) where some of the most stable NIR references are calibrated against some of the most accurate frequency standards, such as Cs fountain clocks, or optical atomic clocks. Using these as frequency references provides the state-of-the-art stabilities and accuracies, but those references have to be made available to remote user laboratories and some spectral transfer from the NIR to the MIR is needed. The former is achieved by using an optical fiber link dedicated to the dissemination of ultra-stable frequency references [24] while an optical frequency comb (OFC) can bridge the gap between the NIR and any region of the MIR. First demonstrations of the stabilization of a QCL on a remote NIR frequency reference have been recently reported [25,26].

In this paper, we report the development of a fully operational widely tunable SI-traceable MIR spectrometer based on a QCL stabilized to a remote NIR ultra-stable reference transferred from the French NMI LNE-SYRTE to Laboratoire de Physique des Lasers (LPL) with an optical fiber link. Compared to our previous work [25], we extend the tuning capability of the set-up and report continuous scanning of the QCL over hundreds of megahertz at a stability improved by more than a factor of 2. We extend the spectral coverage of our spectrometer allowing us to take full advantage of the ~100 GHz QCL's tunability. We couple this cutting-edge MIR spectrometer to a multipass cell which results in high detection sensitivity and allows weak lines to be probed at the low pressures required for ultra-high resolution measurements. We use this novel experiment to carry out ultra-high resolution spectroscopy of relatively high-*J* (29 to 34) ro-vibrational lines of methanol in the C-O stretching mode and demonstrate record uncertainties on resonance frequency measurements. Compiling reliable laboratory spectroscopic data of methanol is important for a variety of reasons. Methanol, the simplest of the organic alcohols, is one of the most abundant interstellar and protostellar molecules and is found throughout the universe in a wide variety of astronomical sources [27]. It is thus an excellent probe of the physical conditions and history of interstellar clouds, but its intrinsically strong spectrum contributes significantly to the "grass" in any interstellar radio astronomy observation and needs to be accurately removed during searches for new molecules. Methanol is important to interstellar chemistry too, because its formation provides a pathway to more complex molecules that are necessary for life . Closer to us, methanol is the second most abundant organic molecules in the Earth's atmosphere after methane [29], it has a significant impact on air quality and is implicated in the production of tropospheric ozone [30]. Retrieving methanol concentration in air from remote-sensing observations is thus a real challenge for environmental and human health issues, and requires accurate laboratory spectroscopic measurements [28]. Although one of the simplest asymmetric-top with a hindered internal rotor, methanol has a rather intricate rotation-torsion-vibration energy structure and is as such also a very important molecule for fundamental infrared and microwave spectroscopy [31–33], metrological applications and frequency calibration [34,35], the realization of optically-pumped far-infrared gas lasers [36] or fundamental physics tests. It has for instance been identified as a very good candidate for probing the limits of the Standard Model because it is one of the most sensitive molecules for a search of a varying proton-to-electron mass ratio [37]. In this work, we focus on very precise measurements of ro-vibrational frequencies of methanol. We carry out saturated absorption spectroscopy from 970 to 973 cm$^{-1}$, thus taking full advantage of the QCL's tunability, we report continuous tuning of the spectrometer over ~400 MHz at the precision of the frequency reference and demonstrate a record sub-10 kHz methanol resonance frequency uncertainty. Spectroscopic measurement at this level of precision in the MIR are scarce, and to our knowledge, there exist only one single other methanol frequency measurement of a weak absorption line around 947.7 cm$^{-1}$ with an uncertainty of 2.4 kHz comparable to ours [34].

## 2. EXPERIMENTAL SET-UP

### 2.1 Widely-tunable QCL stabilized to a remote frequency reference

Fig. 1 shows the experimental set-up and presents the method used to transfer the stability and SI-traceability of a remote NIR frequency reference located at LNE-SYRTE laboratory in Paris to a 10 µm QCL located at LPL laboratory in Villetaneuse (~10 km north of Paris) [38].

The NIR frequency reference signal is given by a 1542-nm ultra-stable laser operated at LNE-SYRTE laboratory. The latter consists of a laser diode locked to an ultra-stable cavity [39,40]. It exhibits a relative frequency stability lower than 10$^{-15}$ between 0.1 s and 10 s. Its absolute frequency is measured using an OFC once per second against primary frequency standards with a total uncertainty of a few 10$^{-14}$ [41]. It is thus referenced to the LNE-SYRTE realization of the SI second. A slow feedback loop, based on this measurement, is used to correct the drift of the ultra-stable cavity. The reference signal is transferred to the LPL laboratory with a noise-compensated optical fiber link. We use a second parallel fiber and the two-way technique [42] to compare the signal frequencies at LNE-SYRTE and that transferred to LPL. We measure instabilities added by the link below 10$^{-15}$ at 1 s averaging time (and even below 10$^{-16}$ when using a Λ-type counter) and a residual frequency bias at a level of 4×10$^{-20}$ [38,43]. The reference signal is thus transferred to LPL without any degradation of its stability and uncertainty.

To allow the QCL to be tuned while maintaining ultimate stabilities and accuracies, we set up at LPL a widely tunable optical local oscillator (OLO). It is based on a local 1.54 µm laser diode emitting at a frequency $\nu_{\text{OLO}}$ which is phase-locked to the NIR reference transmitted from LNE-SYRTE. Tunable sidebands are generated in the OLO signal by phase-modulating the laser diode using a fiber-coupled Electro-Optic Modulator (EOM). The latter is driven by a home-made phase-jump-free microwave synthesizer based on an YIG (Yttrium Iron Garnet) oscillator. The tuning range is currently limited to 7.5-16.5 GHz, as detailed in Section 2.2. As illustrated in Fig. 1, we detect the beat note signal between the upper sideband (at frequency $\nu_{\text{OLO}} + f_{\text{EOM}}$, with $f_{\text{EOM}}$ the microwave frequency at which the EOM is driven) and the NIR reference laser (frequency $\nu_{\text{ref}}$) and process it to phase-lock the OLO sideband onto the NIR reference (using phase-lock loop PLL1 in Fig. 1) with an offset frequency $\delta_1$ given by:

$$\delta_1 = \nu_{\text{ref}} - \nu_{\text{OLO}} - f_{\text{EOM}}. \tag{1}$$

This phase-lock loop allows the OLO frequency to be tuned over 9 GHz by scanning the EOM frequency.

In a second step, one tooth of the LPL OFC, which consists of a $f_{\text{rep}} \approx 250 MHz$ repetition rate erbium-doped fibre mode-locked laser emitting around 1.54 µm, is phase-locked to the OLO carrier frequency with an offset frequency $\delta_2$. To this end, we detect the beat note signal between a comb tooth and the OLO on PD2 and use PLL2 (see Fig. 1) after removal of the comb carrier-envelope offset frequency. The latter is detected via an interferometric *f-2f* self-referencing scheme [44], and is subtracted by mixing it with the beat note signal [45]. This leads to the following equation:

$$\delta_2 = \nu_{\text{OLO}} - N f_{\text{rep}}, \tag{2}$$

with *N*, an integer. This results in phase-locking the comb repetition rate to a subharmonic of the OLO frequency while also transferring the tunability from the OLO carrier to the repetition rate $f_{\text{rep}}$. A bandwidth larger than 500 kHz is obtained by acting on the OFC cavity length with an intra-cavity EOM and a piezo-electric transducer mounted on a cavity mirror.

Finally, a 10.3 µm QCL is phase-locked to the stabilized and tunable OFC (see details in [25,46]). Our QCL of frequency $\nu_{\text{QCL}}$ is a room temperature distributed-feedback source from Alpes Lasers tunable over 170 GHz. To bridge the gap between the NIR and the MIR domains, a custom output of the OFC at 1.81 µm is overlapped with the QCL beam in a AgGaSe$_2$ non-linear crystal for sum-frequency generation. The resulting shifted comb, centered at 1.54 µm (mode frequencies $qf_{\text{rep}} + f_0 + \nu_{\text{QCL}}$, with $q$ an integer and $f_0$ the OFC carrier-envelop offset frequency), is combined with the 1.54 µm main output of the OFC (mode frequencies $pf_{\text{rep}} + f_0$, with $p$ an integer), yielding a beat note signal of frequency $(nf_{\text{rep}} - \nu_{\text{QCL}})$, with $n = p - q$, between the QCL frequency $\nu_{\text{QCL}}$ and the $n^{\text{th}}$ harmonic of $f_{\text{rep}}$ measured on photodiode PD3 (see Fig. 1). This signal is processed (using phase-lock loop PLL3 in Fig. 1) to generate a correction signal applied to the current of the QCL. The QCL is then phase-locked to the $n^{\text{th}}$ harmonic of the OFC repetition rate with an offset frequency:

$$\delta_3 = nf_{\text{rep}} - \nu_{\text{QCL}}. \qquad (3)$$

In a previous work [25], we have demonstrated that this stabilization method allows the excellent stability and frequency uncertainty of the remote ultra-stable NIR reference laser to be transferred to the comb and in turn to the QCL with a bandwidth of around 500 kHz. This results in a QCL stability at the level of the reference, below 0.03 Hz ($1\times10^{-15}$ in relative value) from 1 to 100 s and a line width better than 0.2 Hz for a 100 s measurement time. The remote reference laser frequency being controlled against a combination of a H-maser and a sapphire cryogenic oscillator, themselves monitored on the Cs fountains of LNE-SYRTE, the QCL's frequency is SI-traceable and is known within a total uncertainty better than $4\times10^{-14}$ at 1 s averaging time [41].

Here, we report a crucial advance, the wide tuning capability allowing to scan the QCL frequency at the precision of the reference and to take full advantage of the QCLs' tunability. Scanning the OLO carrier frequency over 9 GHz allows the comb to be widely tuned and in turn the QCL to be continuously tuned over potentially 1.3 GHz (given by the MIR to NIR frequency ratio) while maintaining ultimate accuracies and stabilities. In our current set-up, this tuning range is limited to around 400 MHz due to the piezo-electric actuator currently used to act on the OFC cavity length.

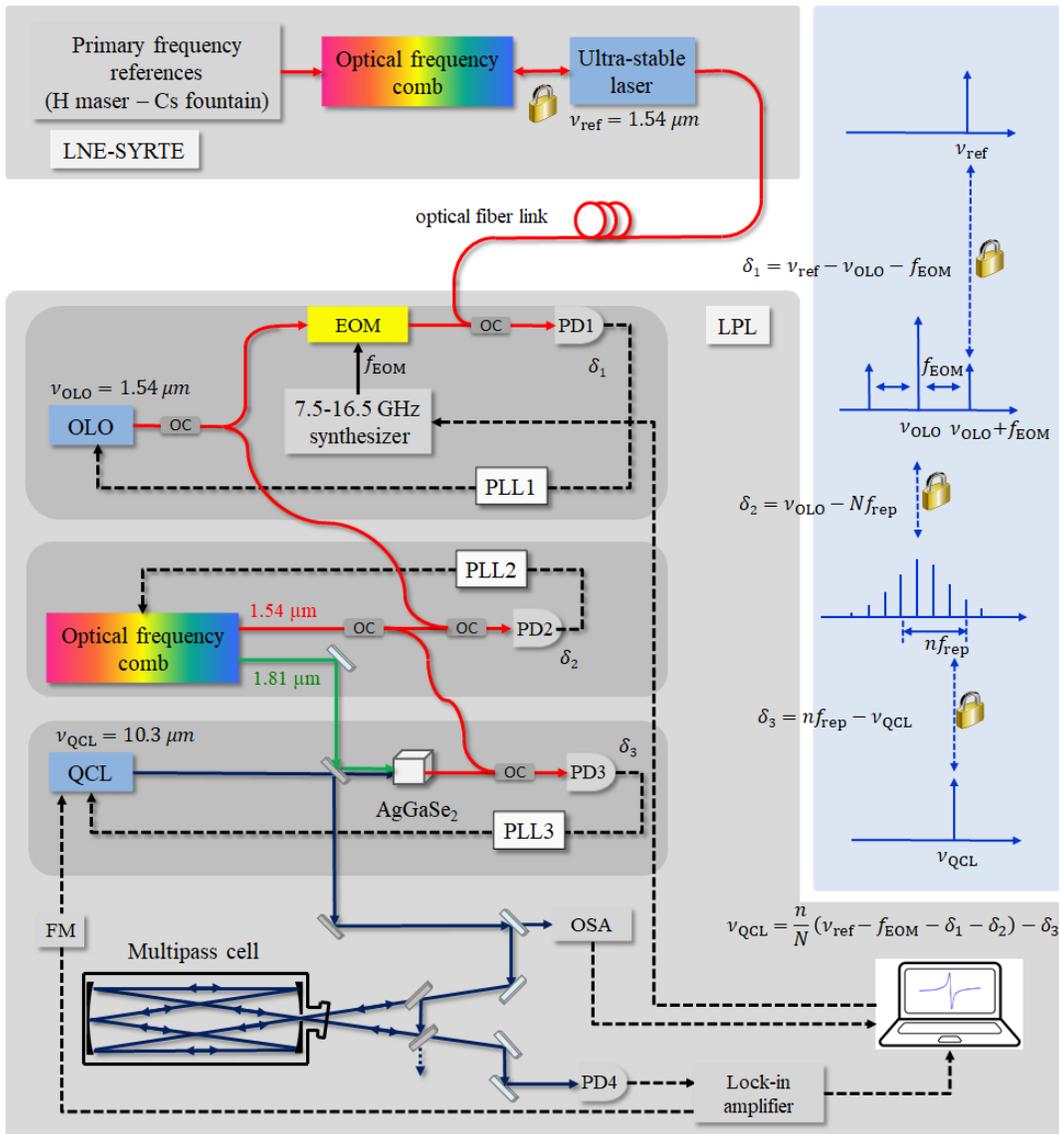

Fig.1 Experimental set-up. The NIR reference signal (frequency $\nu_{\text{ref}}$) developed at LNE-SYRTE (Paris) is transferred to LPL (Villetaneuse) through a 43-km long optical-fiber link with active noise compensation (not shown). At LNE-SYRTE, its absolute frequency is measured against primary frequency standards using an OFC. At LPL, a local laser diode (frequency $\nu_{\text{OLO}}$) is used as an optical local oscillator (OLO). An electro-optic modulator (EOM) driven at frequency $f_{\text{EOM}}$ with a phase-jump-free synthesizer generates sidebands tunable over 9 GHz in the OLO signal. The beat note signal between one sideband and the reference signal provided by LNE-SYRTE is used to phase-lock the former to the latter, with an offset frequency $\delta_1$ (using PLL1). This allows the carrier frequency to be used as a tunable ultra-stable local oscillator. The repetition rate $f_{\text{rep}}$ of an OFC is then phase-locked to the OLO carrier frequency after removal of the comb carrier-envelope offset frequency (*via* PLL2). A QCL (frequency $\nu_{\text{QCL}}$) is finally phase-locked to the stabilized OFC by performing sum frequency generation in a AgGaSe$_2$ crystal and processing the beat note signal between the resulting beam and the OFC (using PLL3). Tunability is thus transferred from the local reference to the OFC and finally the QCL. The stabilized and tunable QCL beam is then used to perform saturated absorption spectroscopy in a multipass cell. The QCL is frequency modulated and the signal is recorded after detection in a lock-in amplifier. PD: photodetector; PLL: phase lock loop; FM: frequency modulation; OC: optical coupler, OSA: optical spectrum analyzer. Padlocks symbolize phase-lock loops.

## 2.2 Home-made phase-jump-free microwave synthesizer

Our stabilization method using phase-lock loops, broad and continuous tunability of the QCL requires the synthesizer driving the EOM to be tuned without phase-jump. As far as we know, phase-jump-free low-phase-noise synthesizers, such as Direct Digital Synthesizers (DDS), are not commercially available in the 8-18 GHz microwave window. We thus developed a home-made phase-jump-free low-phase-noise source based on a YIG oscillator. The latter is phase-locked to a DDS tunable from 0 to ~400 MHz, with a frequency ratio of 64. The synthesizer's tuning range is limited to 7.5-16.5 GHz by the YIG's span. The DDS is referenced to an ultra-stable quartz oscillator disciplined on a GPS (Global Positioning System) receiver. As detailed in Section 3.3.1, it results in a microwave frequency stability below $10^{-12}$ between 1 s and 100 s and a long term frequency uncertainty of $10^{-11}$. When used in our experimental set-up, it contributes at most at the level of $8\times10^{-16}$ to the OLO carrier (and in turn QCL) frequency uncertainty. Our home-made microwave synthesizer is integrated in a rack-mounted package and is computer-controlled. Adjusting the DDS frequency allows the synthesizer frequency $f_{\text{EOM}}$ to be continuously tuned without any phase jump over 9 GHz, limited by the YIG's span. It delivers a power of at least 15 dBm over its entire spectral window.

## 2.3 Spectroscopic set-up

The stabilized QCL is used to record ro-vibrational spectra of methanol around 10.3 μm. We carry out saturated absorption spectroscopy with a simple and quite compact set-up using a multipass cell (see Fig. 1). The laser beam goes through a Faraday isolator and is split using an 80/20 beam splitter. The most powerful beam (~ 10 mW) is sent to the sum frequency generation set-up for locking the QCL to the OFC. The other beam (~2.6 mW) is used for spectroscopy. It is split in a pump and a probe beam which are coupled into an astigmatic Herriott multipass cell (Aerodyne Research, AMAC-36 model, 20 cm base path, 182 paths) with an incident power of the order of 1.5 mW and 0.8 mW respectively. The intensity of the probe beam is detected after transmission through the cell with a liquid-nitrogen-cooled mercury-cadmium-telluride photodetector (PD4 in Fig. 1). The waist of the laser beams at the center of the multipass cell is around 0.8 mm. The effective absorption length of 36.4 m allows us to obtain sufficiently high signal-to-noise ratios at working pressures in the 0.1-10 Pa range, in which range resolutions of a few hundreds of kilohertz can be obtained. A few optical components including the cell are purposely shaken using piezo-electric transducers or fans in order to average out the interference fringes typically observed when using multipass cells [47].

Frequency modulation is used to improve the signal-to-noise ratio (see Fig. 1). Both probe and pump beams are frequency modulated at 20 kHz with an excursion of 50 kHz, much smaller than the typical linewidth of around 700 kHz full-width-at-half-maximum (FWHM). Mainly first-, but also second- and third-harmonic detection is used by demodulating the probe signal in a lock-in amplifier.

## 2.4 Spectra acquisition and absolute frequency determination

Absorption spectra are measured by scanning the frequency $f_{\text{EOM}}$ of the home-made synthesizer driving the EOM and as a result the OLO carrier, the OFC repetition rate and the QCL frequency in a series of discrete steps. The absorption signal is recorded using a data acquisition computer card. Spectra are typically recorded with an EOM frequency step (and thus an OLO carrier frequency step) $\Delta f_{\text{EOM}} = 100$ kHz and a step duration of 100 ms. This results in steps of ~125 mHz and ~15 kHz for $f_{\text{rep}}$ and $\nu_{\text{QCL}}$ respectively. The time constant of the lock-in amplifier is also chosen to be 100 ms (which allows us to obtain ample signal-to-noise ratio in a single scan). During the scan, the comb repetition rate $f_{\text{rep}}$ is recorded in real time. The reference absolute frequency $\nu_{ref}$ measured once per second at LNE-SYRTE is saved in real time on a repository accessible *via* hyper-text transfer protocol (http). The Network Time Protocol (NTP) is used for time synchronization between the various computers used at LPL and LNE-SYRTE. We also monitor the QCL frequency using an optical spectrum analyzer (OSA in Fig. 1, Bristol 771B, 1 ppm frequency accuracy).

Let us denote $f_{EOM}^0$, $f_{rep}^0$ and $\nu_{QCL}^0$ the EOM frequency, the comb repetition rate, and the QCL frequency at the start of a scan, the latter measured using the optical spectrum analyzer. According to equation (1), the corresponding OLO frequency is:

$$\nu_{OLO}^0 = \nu_{ref} - f_{EOM}^0 - \delta_1. \tag{4}$$

The two integers $N$ and $n$ are then unequivocally determined using equations (2) and (3) and rounding to the nearest integer (square brackets):

$$N = \left[\frac{\nu_{OLO}^0 - \delta_2}{f_{rep}^0}\right] \tag{5}$$

$$n = \left[\frac{\nu_{QCL}^0 + \delta_3}{f_{rep}^0}\right], \tag{6}$$

Once $N$ and $n$ determined, the absolute frequency of the QCL is calculated using again equations (2) and (3):

$$\nu_{QCL} = \frac{n}{N}(\nu_{ref} - f_{EOM} - \delta_1 - \delta_2) - \delta_3 . \tag{7}$$

Note finally that all beat notes are counted using dead-time free counters (K+K FXE) operated in Π-type mode with a gate time of 1 s [48].

We have also developed an alternative protocol for retrieving the absolute frequency scale which does not require to know the optical reference absolute frequency $\nu_{ref}$. Instead, it uses an ultra-stable radio-frequency reference disciplined to GPS. This alternative method is described in the Appendix. We have systematically processed our data with both protocols for cross-checking and to identify potential sources of error.

## 3. PRECISE SPECTROSCOPY OF METHANOL

We have recorded saturated absorption spectra of about twenty lines of methanol around 29.1 THz (between 970 cm$^{-1}$ and 973 cm$^{-1}$). Most of them belong to the $P$ branch of the $\nu_8$ C-O stretch vibrational mode [49], with $J$ (the total orbital angular momentum quantum number) ranging from 29 to 34. At a typical pressure of 1 Pa, we measure an absorption of about 30 % for the most intense lines. For high-resolution saturated absorption experiments, we carry out frequency scans in both sweep directions, with increasing and decreasing frequencies and consider only averages of a pair of such up and down scans. In this way, frequency shifts induced by the limited detection bandwidth are eliminated [50]. We typically record five such pairs spanning ~15 MHz in the MIR (corresponding to an EOM frequency scan of ~100 MHz). Fig. 2 shows a saturated absorption spectrum of the P($E$,co,0,2,33) ro-vibrational transition (see Appendix for details on the spectroscopic notations) averaged over 5 pairs, after retrieval of the absolute frequency scale following the procedure of Section 2.4. It exhibits a signal-to-noise ratio of ~650 and a peak-to-peak linewidth of ~400 kHz (~700 kHz FWHM), which is a combination of the pressure broadening (previously measured to be between 100 and 300 kHz/Pa FWHM [34,51]), transit time broadening ($2u_m/8w_0$ ~120 kHz FWHM [52], with $u_m$ the most probable velocity and $w_0$ the beam waist) and power broadening.

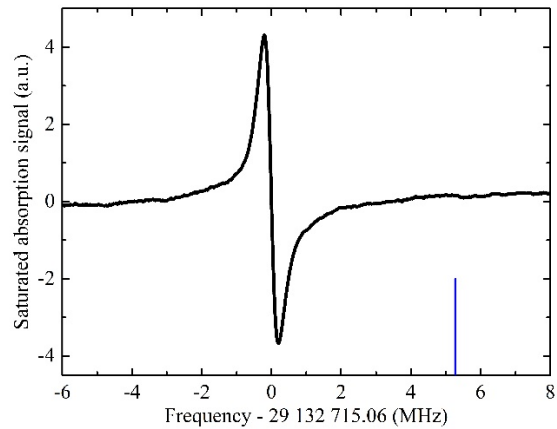

Fig.2 Saturated absorption spectrum of the P($E$,co,0,2,33) ro-vibrational line of methanol recorded using frequency modulation and first-harmonic detection. Experimental conditions: pressure: 1 Pa; modulation frequency: 20 kHz; frequency modulation excursion: 50 kHz; frequency step: ~15 kHz; average of 5 pairs of up and down scans; total integration time per point: 1 s; whole spectrum measurement time: 935 s. The P($E$,co,0,2,33) line frequency position reported in the HITRAN database [53] at 971.76295 cm$^{-1}$ is also shown as a blue stick.

*3.1 Tunability and spectral coverage*

Fig. 3 demonstrates the unprecedented continuous tuning capability resulting from the implementation of our widely tunable frequency stabilized OLO. It shows a spectrum spanning 400 MHz and exhibiting 5 neighboring lines assigned using the notations detailed in the Appendix. The oscillations in the baseline result from the Doppler broadening contribution to the line shape. The grey solid line is a fit to the data. Each line is fitted with the first derivative of a sum of a Lorentzian and a Gaussian to model the saturated absorption and Doppler contribution respectively. The Gaussians' FWHM is found to be 64 MHz, in perfect agreement with the expected Doppler broadening. This continuous tuning range of 400 MHz is a 4-fold improvement on our previous work [25]. In the latter, the QCL frequency was swept by scanning the frequency of the synthesizer used to phase-lock the QCL onto the OFC, and so the scanning window was limited to about half of the comb repetition rate, *i.e.* ~100 MHz. The tuning range of the stabilized QCL is currently around 400 MHz corresponding to 3 kHz tuning range of the comb repetition rate. The latter is limited by the piezo-electric actuator used to move a cavity mirror. The QCL tuning range could be increased to 1.3 GHz (a limit set by our current home-made microwave synthesizer taking into account the MIR to NIR frequency ratio) by using a step motor or by tuning the temperature of the comb.

The tuning capability is also achievable anywhere in the entire QCL spectral range allowing us to take full advantage of the QCL's tunability. Fig. 4 shows that we have been able to record about 20 lines (represented as dashed lines) in a spectral window covering ~90 GHz, *i.e.* anywhere where sufficient power is available for locking the QCL to the OFC. In particular, this allows us to access regions inaccessible to frequency-stabilized $CO_2$ lasers typically used for precise spectroscopic measurements in this region. Note also that the spectral coverage of our set-up can be extended to the entire 9-11 µm range without any modification of the comb or non-linear crystal but using a series of QCLs of adjacent emission spectrum. In the following, we will focus on the P(*E*,co,0,2,33) ro-vibrational line shown in Fig 2 and pinpointed with an asterisk in Fig. 4 (around 971.75 $cm^{-1}$), in order to assess our sensitivity in measuring absolute frequencies of methanol ro-vibrational lines.

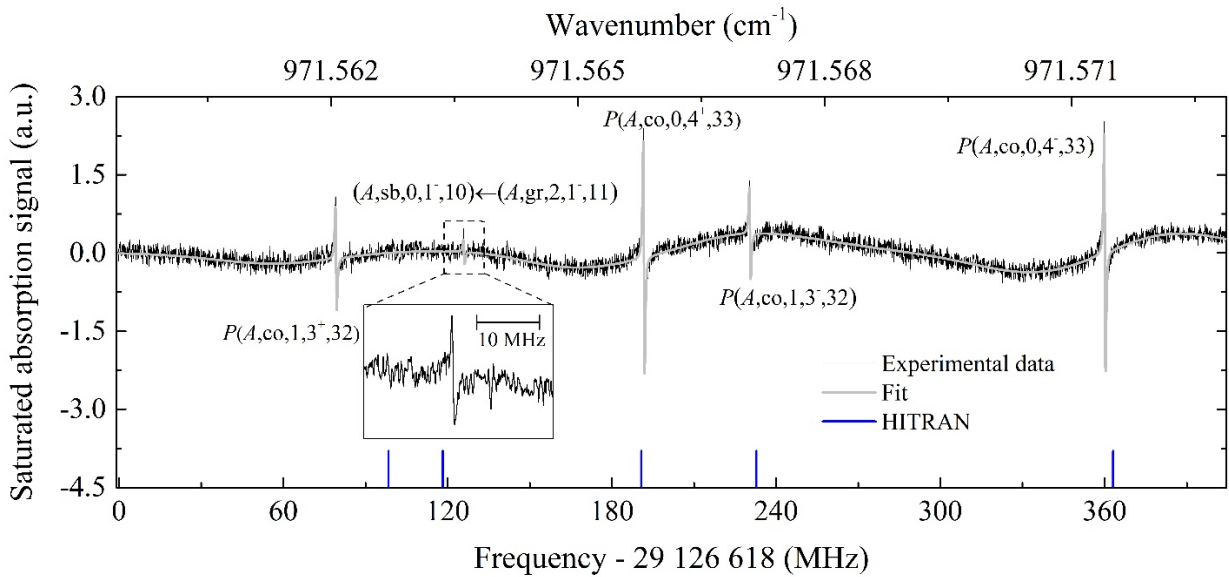

Fig.3 Saturated absorption spectrum of methanol spanning 400 MHz recorded using frequency modulation and first-harmonic detection (black line). The grey solid line is a fit to the data. The oscillations in the baseline result from the Doppler broadening. The observed lines are labeled with the transition assignments as detailed in the Appendix. Blue sticks indicate the line frequency positions reported in the HITRAN database [53]. The inset is a zoom on the small (*A*,sb,0,1⁻,10)←(*A*,gr,2,1⁻,11). Experimental conditions: pressure: 1 Pa; modulation frequency: 20 kHz; frequency modulation excursion; 50 kHz; frequency step: ~15 kHz; average of 1 pair of up and down scans; total integration time per point: 200 ms; whole spectrum measurement time: 5400 s.

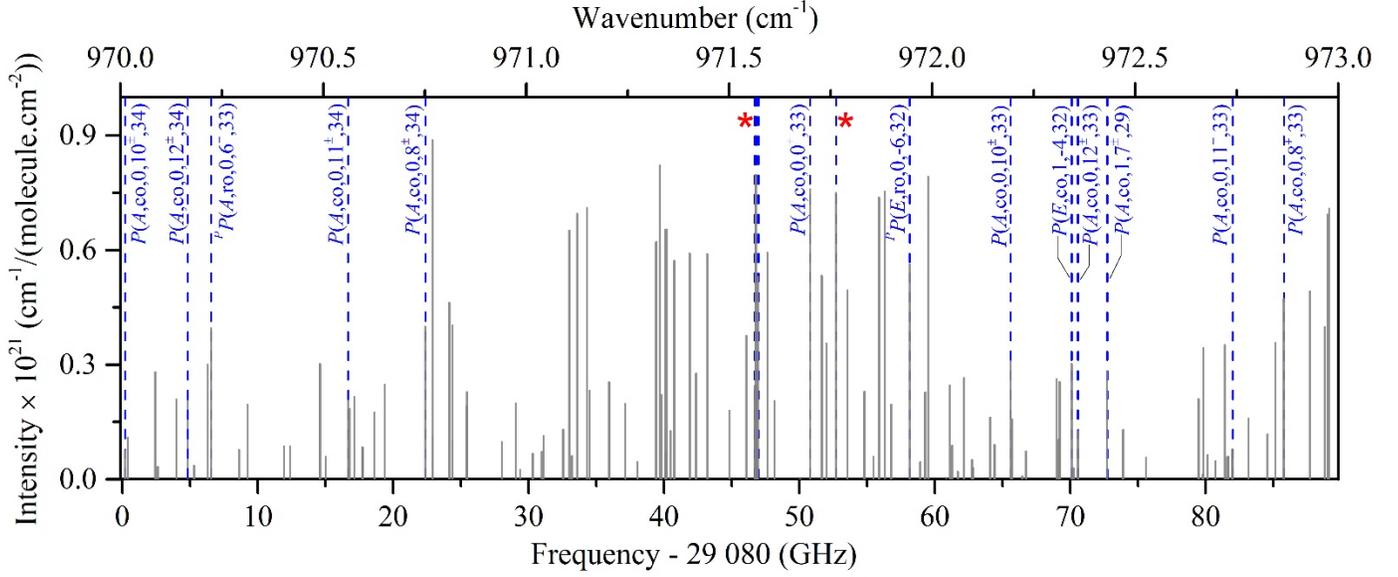

Fig.4 Spectral range coverage. Grey solid sticks: spectrum of methanol simulated using HITRAN's data [53]. Blue dashed sticks: methanol lines recorded by us so far labeled with their spectroscopic assignment (see Appendix). The spectral range covered is of ~90 GHz. The red asterisks pinpoint the data shown in Fig. 2 (thin stick around 971.75 cm$^{-1}$) and 3 (thick stick around 971.56 cm$^{-1}$) above where their assignments have been introduced.

### *3.2 Methanol frequency measurements*

The saturated absorption line shape is expected to be rather intricate when using a multipass cell. The mirrors' reflectivity results in a 15 to 20% transmission after 182 passes. The pump and probe powers thus inversely vary by almost an order of magnitude through the cell, leading to a pump-to-probe power ratio ranging from ~15 to ~0.3, a ~50% total power variation depending on the pass number, and in turn to marked fluctuations of the saturation parameter. The resulting line shape is thus likely to deviate from the Lorentzian profile typically used in the pressure broadened regime of our experiments. Besides, as seen in Fig. 2, the observed line shape exhibits an asymmetry. Several effects associated with line broadening can contribute to the line shape asymmetry, amongst which the residual amplitude modulation associated with frequency modulation, the gain-curve-induced laser power variation over a frequency scan, and the gas lens effect-induced distortion (see Section 3.3.2) are likely causes. Those line shape issues ultimately limit the determination of the transition center frequency (see Section 3.3). We adopt a phenomenological approach to deal with our lack of knowledge of the correct model for the line shape by fitting to the data different profiles, including various sums of derivatives of increasing orders (both odd and even) of a Lorentzian, Voigt profiles and/or a dispersion line shape, considering or not the Beer law, as well as polynomial baselines of various orders. We also vary the spectral range over which the fit was performed from 15 MHz to 6 MHz. The model that best matches the data is found to be:

$$\frac{\partial}{\partial \nu} g(\nu) + \sum_{i=0}^{N} s_i (\nu - \nu_0)^i,$$
$$\text{with } g(\nu) = A[B(\nu - \nu_0) + 1] \times L(\nu) \text{ and } L(\nu) = \frac{1}{\pi} \frac{\gamma}{(\nu - \nu_0)^2 + \gamma^2}, \qquad (8)$$

$A$, $\gamma$ and $\nu_0$ represent respectively the area, the half-width-at-half-maximum and the center of the Lorentzian curve $L(\nu)$, $B(\nu - \nu_0)$ factors in the asymmetry and $\sum_{i=0}^{N} s_i (\nu - \nu_0)^i$ is a polynomial of order N which models the baseline. Note that $g(\nu)$ corresponds to the sum of a Lorentzian and a dispersion shape. A seventh order polynomial is needed in order to correctly fit the spectrum. As can be seen in Fig. 2, the baseline exhibits some residual, non-stationary, interference fringes, which we attribute to the limited efficiency of our strategy to scramble multipass cell-induced etaloning effects (see Section 2.3). Besides fitting those fringes, the polynomial baseline may also well catch some non-Lorentzian contribution to the experimental profile.

We also carried out measurements using second and third harmonic detection to which we also fitted the variety of line shape profiles listed above. Note however that the baseline being strongly flattened in these cases, it was no longer needed to include it in the model. Amongst all the models considered, $g(\nu)$ is the simplest for which fitting to first, second and third harmonic detection data (using the corresponding derivatives of $g(\nu)$) all converged to a transition center frequency to within 5 kHz. Most of the other profiles happen to be inconsistent at the few tens of kilohertz level, which gives $g(\nu)$ the necessary credit, further confirmed by the exemplary outcome of our least-squares procedure (chi-square values and residuals).

As mentioned above, a single measurement typically consists in recording 5 pairs of up and down spectra. We then perform a 'pair by pair' analysis. We average each pair and assign the same error bar to all data points of such a single averaged spectrum. This experimental error bar is calculated as the standard deviation of the residuals obtained after fitting a second-order polynomial to a small portion of the spectrum far from resonance. It is typically of the order of 1% of the signal peak-to-peak amplitude. Fig. 5 shows one such averaged pair of saturated absorption spectra of the P($E$,co,0,2,33) ro-vibrational transition of methanol (corresponding to one of the 5 pairs contributing to Fig. 2), the resulting fit to the data and its residuals. The resulting uncertainty on the central frequency is around 1 kHz, in good agreement with a simple estimate consisting in multiplying the experimental error bar and the signal slope at the center of the resonance. However, we find that the dispersion of the five frequencies resulting from the analysis of the 5 pairs is larger than the uncertainty resulting from the fit to a single pair. We then estimate the transition absolute frequency and the associated uncertainty for one measurement by calculating the weighted mean of the five fitted absorption frequencies (we use the fits' error bars to calculate weights) and the corresponding weighted standard error. The latter ranges from 0.6 to 3.9 kHz for all our measurements.

We have carried out about 20 such measurements, in June and October 2017. In June, we used a fiber link without any active noise compensation to transfer the optical reference from LNE-SYRTE to LPL, while in October we used the active-noise compensated fiber link. Between June and October, a better fringe scrambling system was installed, resulting in a signal-to-noise ratio increased by a factor of about 1.5. All measurements were carried out at a pressure of 1 Pa, but the intra-cell average laser power used for spectroscopy was reduced by about 30%, from 1.1 mW in June to 0.8 mW in October. We define the intra-cell average laser power as the total power from the two counter-propagating beams averaged over the 182 paths inside the cell. Fig. 6 displays the transition frequencies and their uncertainties for 20 measurements of the P($E$,co,0,2,33) ro-vibrational methanol line previously shown in Fig. 2 and Fig. 5, after correcting for a power-dependent frequency shift (see details in Section 3.3.2).

The transition frequency resulting from the weighted mean of all data extrapolated at zero-power is 29 132 715 076.7 (1.0) kHz. The transition frequency resulting from the weighted mean of all June – respectively October – data extrapolated at zero-power shown in Fig. 6 is 29 132 715 075.3 (1.3) kHz – respectively 29 132 715 077.5 (0.7) kHz. The 1-σ weighted standard errors quoted into parentheses correspond to our statistical uncertainties. The 2.2 kHz difference between the data recorded in June and October is compatible with the combination of the 1.4 kHz uncertainty resulting from the power-dependent shift correction and the statistical uncertainty (see below). The notable progress in measurement repeatability between June and October (shown by the corresponding improvement in weighted standard deviations from 3.5 kHz to 1.3 kHz) results from the improvements made to the experimental set-up between the two runs and to the resulting increase in control and stability.

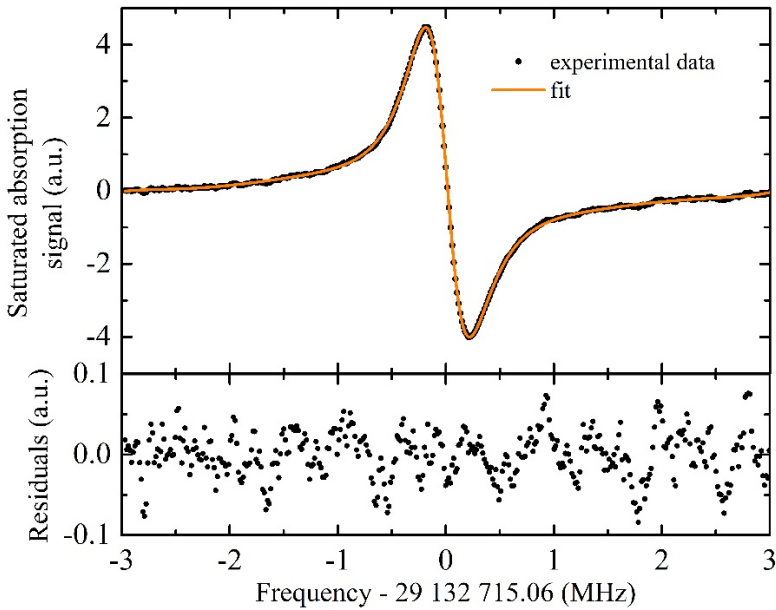

Fig.5 Saturated absorption spectrum of the P($E$,co,0,2,33) ro-vibrational line of methanol, with fit and residuals. The data (black dots) are recorded using frequency modulation and first-harmonic detection. The orange solid line is a fit to the data resulting in a reduced chi-squared of 1.27. Residuals are shown on the bottom graph. Experimental conditions: pressure: 1 Pa; modulation frequency: 20 kHz; frequency modulation excursion; 50 kHz; frequency step: ~15 kHz; average of 1 pair of up and down scans (corresponding to one of the 5 pairs contributing to Fig. 2); total integration time per point: 200 ms; whole spectrum measurement time: 80.2 s.

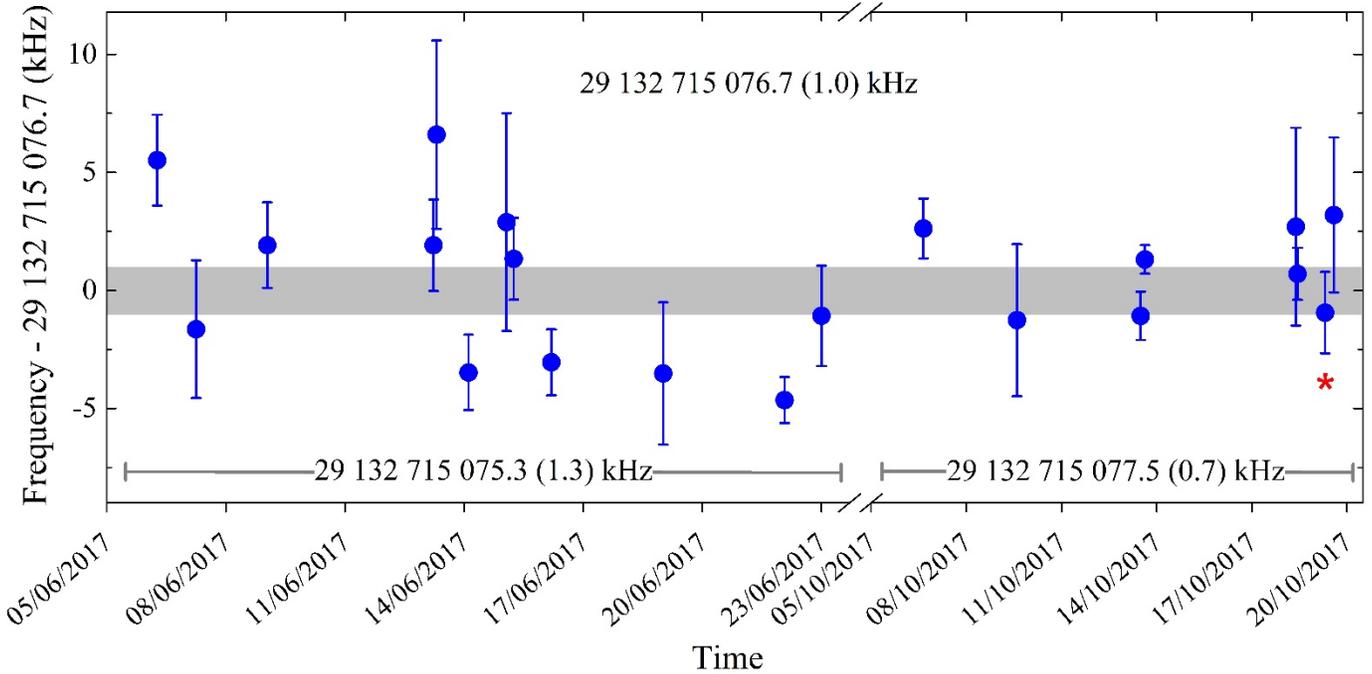

Fig.6 Frequency measurements of the P(*E*,co,0,2,33) ro-vibrational methanol line. Twenty measurements carried out at a pressure of 1 Pa in June and October 2017 are displayed after correction of a power-dependent frequency shift. The transition frequency resulting from the weighted mean of all June (respectively October) data are indicated at the bottom with the associated 1 σ weighted standard errors. The overall weighted mean and its 1 σ weighted standard error is indicated at the top and symbolized by the grey rectangle. The red asterisk pinpoints the measurement corresponding to Fig. 2 and 5.

*3.3 Frequency measurement uncertainty budget*

3.3.1 Frequency scale

The uncertainty of the frequency scale is limited by both the optical reference and the frequency tuning system. The frequency of the former is measured at LNE-SYRTE against an H-maser itself monitored against primary standards [41]. As mentioned above, its absolute frequency is known with an uncertainty better than $4\times10^{-14}$. The uncertainty associated to the transfer link is better than $10^{-19}$ when the propagation-induced phase noise is actively compensated [38,43]. When free-running, the link mean frequency offset is below $10^{-14}$ and so is the link instability for averaging times longer than a few seconds. We take this as a conservative upper limit on the transfer uncertainty, which happens to be negligible at the level of our molecular frequency measurement uncertainties (see below). Note that the noise of the fiber link, whether it be free-running or compensated was monitored during each of our frequency measurements.

At LPL, this reference frequency is locally disseminated to the experimental set-up using a 10-m-long optical fiber without any passive or active noise compensation. Taking into account air conditioning-induced ~1-K peak-to-peak temperature fluctuations, we estimate the resulting free-running fiber phase noise to lead to a frequency instability below $10^{-15}$ at 1 s and decreasing at longer averaging times with a bump at ~$10^{-15}$ for a measurement time of a few hundreds of seconds corresponding to the air conditioning half cycling period [54].

The reference frequency is sampled every second at LNE-SYRTE. The synchronization between the LNE-SYRTE and LPL measurement times is estimated with an uncertainty below 100 ms by using an NTP connection to the LNE-SYRTE server. The 1-10 Hz/s mean drift of the ultra-stable laser being removed (see Section 2.1), the residual frequency drift lies below $10^{-2}$ Hz/s. The imperfect synchronization between the two laboratories has thus no impact on the frequency uncertainty.

The traceability of the QCL's frequency to the LNE-SYRTE primary standards is ensured by using phase-lock loops for frequency stabilizing the OLO, the comb and the QCL. All involved counters and synthesizers including the DDS used for the YIG oscillator frequency control are referenced to the same local radio-frequency oscillator. It consists of an ultra-stable quartz oscillator disciplined on a GPS receiver with a time constant around 5000 to 10000 s. The frequency of this oscillator is also measured against a radio-frequency reference disseminated from LNE-SYRTE *via* the optical link and calibrated there to the H-maser, itself monitored against primary standards. This shows that the GPS disciplined quartz oscillator has a stability better than $10^{-12}$ between 1 and 100 s and a long-term frequency stability better than $10^{-11}$. Such a measurement ensures the traceability of our local radio-frequency oscillator to primary standards, and allows us to determine its frequency to better than $10^{-11}$. This results in an uncertainty on the OLO carrier frequency and, in turn, on the MIR frequency scale, of the order of $8\times10^{-16}$, limited by the YIG oscillator's frequency uncertainty.

For a typical measurement consisting of 5 pairs of up and down scans, the uncertainty on the MIR frequency scale is thus dominated by the frequency uncertainty of $4\times10^{-14}$ of the LNE-SYRTE ultra-stable reference. This has a negligible effect on our molecular resonance frequency measurements (see below).

### 3.3.2 Spectroscopic effects

A significant shift of the line center is observed when varying the laser power. This power-dependent or ac-Stark shift was investigated by recording spectra for intra-cell average laser powers ranging from 0.05 mW to 1.1 mW. This allowed the determination of a power-dependent shift coefficient of -15.4 (1.3) kHz/mW and the extrapolation to a zero power transition frequency of each June and October measurements by applying a correction of +16.94 kHz and +12.32 kHz respectively. Note that we estimate an upper bound in power fluctuations of 10% during a measurement and from one to another. This leads to frequency fluctuations of the order of 1 kHz which contribute to our statistical uncertainty. However, the 5% specified accuracy of the power meter results in a systematic uncertainty of at most 0.75 kHz at our highest power of 1.1 mW, but doesn't affect our extrapolation to zero power.

The P($E$,co,0,2,33) transition frequency was measured at different pressures in the range 0.1–9 Pa. This allowed the determination of a pressure-dependent shift coefficient of +2.4 (1.0) kHz/Pa. This is of similar magnitude but opposite in sign compared to the only, to our knowledge, other MIR measurement available in the literature carried out on a different methanol transition [34]. It can be used to deduce the collisionless transition frequency. For the measurements in Fig. 6 recorded at a pressure of 1 Pa, the uncertainty associated to this pressure shift correction is 1.0 kHz. Pressure fluctuations smaller than 0.02 Pa during a typical 1000 s measurement (which factor in the leak/outgassing rate of the multipass cell measured at the level of $10^{-5}$ Pa/s) and from one to another result in frequency fluctuations smaller than 0.1 kHz which contribute to our statistical uncertainty at a negligible level. By measuring pressures using different gauges of various technologies, we estimate an upper bound on the pressure measurement accuracy to be 0.02 Pa which translates into a 0.05 kHz frequency inaccuracy. The overall pressure shift correction uncertainty is thus 1.0 kHz.

No wavefront-curvature-induced residual Doppler shift is expected provided that the beam waist is well centered in the multipass cell. This is however only achieved to within a few centimeters resulting in a potential systematic shift smaller than 10 kHz [52,55] (*i.e.* 8% of the transit-time broadening) associated to each of the 182 individual paths of the cell. This curvature-induced shift can be negative or positive depending on which side of the cavity center the beam waist is located. In our astigmatic Heriott multipass cell, from path to path, the beam waist will alternatively be displaced nearly symmetrically on one side of the cavity center and the other. Moreover, no particular care was taken to match the two counter-propagating beams which thus most probably focus at slightly different positions. For these reasons, we expect the overall systematic shift resulting from the combination of the 182 path to average out to less than 1 kHz. Finally, due to regular realignments of the optical set-up, the residual curvature-induced shift fluctuates from one measurement in Fig. 6 to the other in a range smaller than ±1 kHz, and therefore simply contributes to our statistical uncertainty. Gas-lens effects stem from the transverse dependence of the medium index of refraction that results from the interaction with a Gaussian transverse profile laser beam. The medium then behaves as a converging or diverging lens, depending on which side of the resonance the laser is tuned which leads to a beam diameter decrease or increase. Any clipping optics (in our case the central coupling hole of the multipass cell front mirror through which the beam also exits the cell) will then lead to a frequency-dependent power variation of dispersive nature which distorts the line shape [21,56]. We estimate a conservative upper bound of the resulting frequency shift to be +5 kHz [55,57] and take this as our uncertainty. Saturation spectroscopy leads to a photon recoil doublet that is symmetric to the recoil-free transition center and does not produce any shift [58]. For methanol at ~970 cm$^{-1}$, the doublet splitting is ~120 Hz. The second-order Doppler shift is calculated to be ~25 Hz at the most probable methanol velocity at room temperature, with negligible fluctuations given a conservative upper limit 2 K temperature variation [59]. The magnetic frequency shift due to the Zeeman effect is expected to be of the order of a few tens of Hz, with variations 100 times smaller [21]. The black-body radiation shift of ro-vibrational molecular lines has been calculated for a few simple molecules only [60] but it has never been observed. It is expected to be weak with negligible fluctuations within a 2 K temperature variation range. The detailed study of the effects listed in this last paragraph is beyond the scope of this paper. Their contributions are included in Table 1 below under 'other spectroscopic effects'.

### 3.3.3 Line fitting

As mentioned above, by fitting derivatives of $g(\nu)$ to first, second and third harmonic detection data, all fits converged to a transition center frequency within 5 kHz. We therefore assign a conservative 5 kHz systematic uncertainty to the lack of knowledge of the correct model for the line shape. This ultimately limits the determination of the transition center frequency (see Table 1) and covers the multipass cell-induced profile complication, the asymmetry issue and the difficulties resulting from interference fringes affecting the baseline.

The modulation frequency (20 kHz) and frequency excursion (50 kHz) used are both much smaller than the line width (~700 kHz FWHM around 1 Pa). We thus expect negligible frequency modulation-induced distortion of the line shape. We have investigated this by recording spectra using different frequency excursions ranging from 25 kHz to 250 kHz and using first or third harmonic detection. The fitted center frequency did not show any evidence of dependence upon the frequency excursion within our statistical uncertainty.

3.3.4 Budget table and discussion

Table 1 gives the global uncertainty budget. In particular, it summarizes all listed systematic effects, the corresponding corrections to apply to the measured resonance frequency and the associated uncertainties. We finally deduce a zero-power and collision-free transition frequency of 29 132 715 074.3 (7.4) kHz for the P($E$,co,0,2,33) resonance frequency. The $3 \times 10^{-10}$ uncertainty is a factor of 2000 improvement over previous measurements performed using Fourier transform infrared (FTIR) spectroscopy [49] which have led to the creation of an empirical line list used in the current edition of the HITRAN database [53]. Our measured P($E$,co,0,2,33) resonance frequency is shifted by -5263,1 kHz with respect to this previous measurement carried out at a pressure of 13 Pa and reported in the HITRAN database (see also Fig. 2 above). This shift is consistent with deviations observed between previously measured $CH_3OH$ saturated absorption spectroscopy data [61–63] and the FTIR data contributing to the HITRAN database [49,64]. This discrepancy has been attributed to calibration imperfections of FTIR spectrometers as discussed in [35,49,61]. Note that similar shifts are also observed in Fig. 3 for other lines located in a more congested region and will be presented elsewhere.

Other saturated absorption spectroscopy measurements in the C-O stretch vibrational band of methanol are found in the literature. The frequencies of almost 700 transitions have for instance been measured from 1016 to 1063 $cm^{-1}$ with an accuracy of the order of 100 kHz using a $CO_2$-laser–microwave-sideband spectrometer [35,61–63]. Most of those however belong to the $Q$ and $R$ branches, with a small fraction only of low $J$ lines in the $P$ branch. There is, to our knowledge, only one single other frequency measurement of a weak absorption line around 947.7 $cm^{-1}$ with an uncertainty of 2.4 kHz comparable to ours, which was carried out using a heterodyne $CO_2$ laser spectrometer [34].

**Tab.1 Uncertainty budget table for frequency measurements in June and October 2017, all recorded at 1 Pa and displayed in Fig. 6.**

| Systematics | Correction (kHz) | Uncertainty (kHz) |
| --- | --- | --- |
| frequency calibration | 0 | 0.001 |
| power shift | +16.94 kHz for June measurements<br>+12.32 kHz for October measurements | 1.4 |
| pressure shift | -2.4 | 1 |
| other spectroscopic effects | not measured<br>estimated to be <5 kHz | 5 |
| line fitting | 0 | 5 |
| **Total systematics** | +14.54 for June measurements<br>+9.92 for October measurements | 7.3 |
| **Statistics** | 0 | 1.0 |
| **Total** | +14.54 for June measurements<br>+9.92 for October measurements | 7.4 |

## 4. CONCLUSIONS AND PERSPECTIVES

We have developed an apparatus for the precise and tunable frequency control of MIR QCLs over a wide spectral region. A QCL phase-locked to a 1.5 µm optical frequency comb can be continuously scanned at the precision of an ultra-stable frequency reference remotely transferred from an NMI with traceability to primary frequency standards. We have demonstrated continuous tunability over a span of ~400 MHz, a 4-fold improvement compared to previous measurements at such ultra-high levels of spectral purity [25,26]. This continuous span range could be further increased by a factor 3 with simple improvements of the tuning capabilities of the frequency comb. We have used the apparatus to carry out saturated absorption spectroscopy of ro-vibrational lines of methanol in the $P$ branch of the C-O stretching mode in a spectral window of ~90 GHz corresponding to ~50% of the QCL's tuning range. This allows methanol ro-vibrational frequencies to be determined with a ~1 kHz statistical uncertainty and a sub-10 kHz global uncertainty, a ~2000 improvement over previous measurements. The use of a multipass cell provides high detection sensitivity allowing relatively high-$J$ (around 30) ro-vibrational lines to be probed at the low pressures required for ultra-high resolution measurements. This unique apparatus provides an unprecedented combination of resolution, tunability, and frequency and detection sensitivity of any MIR spectrometer to date. Moreover, by using appropriate non-linear crystals and shaping the frequency comb spectrum, our stabilization method currently limited to the 9-11 µm window is extendable to the entire 5-20 µm molecular fingerprint region. This work is also part of a global effort towards ever more precise ro-vibrational spectroscopic measurements in different spectral regions using various frequency controlled laser sources (see for instance the recent studies [16,18–20,25,26,46,65–69]).

Importantly, our method hinges on a fully monitored SI-traceable ultra-stable frequency dissemination facility exhibiting instabilities below $10^{-15}$ at 1 s averaging time. This allows the reference laser ultra-low frequency instability to be transferred to the QCL and our measurement absolute frequency scale to be controlled with an uncertainty at the hertz level, thanks to the traceability to the primary frequency standards developed at LNE-SYRTE [41]. This is a factor of 250 better than GPS-disciplined radio-frequency reference based measurements [15,17–20,67,68] and an improvement of more than six orders of magnitude compared to using traditional frequency synthesizers. These two methods in addition are not SI-traceable. Although at present only a few laboratories have a working optical link to a NMI delivering ultra-stable references, serious efforts are currently being carried out to extend such facilities in France and at the European scale [70,71]. Our method results in QCL's frequency instabilities below $10^{-15}$ for averaging times longer than 1 s and in a record linewidth smaller than 0.2 Hz [25]. This level of spectral purity is important for pushing back the limits in ultra-high resolution molecular spectroscopy. Combining it in a Ramsey interferometry measurement with one of the recently demonstrated cold and slow-moving molecular source technologies [72–76] would give record-breaking tens of hertz or smaller fringe spacings, more than an order of magnitude improvement over the previous record obtained at LPL [5]. Such experiments could potentially lead to molecular frequency measurements at uncertainties reaching the $2 \times 10^{-16}$ ultimate accuracy of the Cs fountains [41], or even better by controlling the frequency of the remote NIR reference with new generation optical clocks [66,77].

Although not at the latter extreme levels, the results reported in this work are at a degree of resolution requiring us to evaluate subtle systematic effects such as the wavefront-curvature and gas-lens effect-induced frequency shift usually ignored. Note that these are magnified when using a multipass cell, which in addition leads to etaloning effects and the resulting difficulties in the line shape modelling. The two main contributions to the uncertainty budget in Table 1 can thus be reduced by replacing the multipass cell by a Fabry-Perot cavity a few meter long, which also results in an improved resolution [25]. With such a cavity, molecular resonance frequency uncertainties as low as a few tens of hertz are within reach [25], but this is at the expense of the tunability which is then limited by the ~100 MHz free spectral range. The multipass cell configuration is thus particularly appropriate for broadband investigations at a high level of resolution while the Fabry-Perot cavity is needed for the most precise measurements.

The particular combination of broad tunability, high sensitivity and resolution demonstrated in this article will participate in driving the next generation spectroscopic technology. It is in particular crucial for future precise spectroscopic tests of fundamental physics, whether it be to provide stringent tests of quantum electrodynamics or precise measurements of the electron-to-proton mass ratio ($m_e/m_p$) [78–80] or to search for physics beyond the standard model by measuring possible variations of $m_e/m_p$ [5,81,82], and looking for fifth forces or extra dimensions at the molecular scale [79,80]. At LPL, we will in particular make the most of this new equipment to pursue our experimental investigations of the frequency differences between the two enantiomers of a chiral molecule induced by the parity violation inherent in the weak interaction [7,72]. Beyond fundamental physics, the proposed technology will facilitate a wealth of precise spectroscopic measurements of interest for space and atmospheric physics and chemistry to improve our understanding of various environments including planetary atmospheres and interstellar gas clouds.

## APPENDIX

*1 Alternative protocol using a radio-frequency reference for frequency scale calibration to primary standards*

We have also developed an alternative protocol for retrieving the absolute frequency scale which does not require to know the absolute frequency of the LNE-SYRTE optical reference $\nu_{\text{ref}}$ but uses a local radio-requency reference which is calibrated to primary frequency standards. It consists of an ultra-stable local quartz oscillator disciplined to the GPS and monitored with an SI-traceable radio-frequency reference (see Section 3.3.1). Here, $n$ is again unequivocally determined using equation (6), but we then simply deduce the QCL absolute frequency scale from equation (3):
$$\nu_{\text{QCL}} = nf_{\text{rep}} - \delta_3 \, ,$$
using the measurement of $f_{\text{rep}}$ carried out with a counter referenced to the local quartz oscillator. The relative uncertainty on the measurement of $f_{\text{rep}}$ is at the level of $10^{-11}$, limited by the uncertainty of the local radio-frequency reference (see Section 3.3.1). And so this alternative method results in an uncertainty on the MIR frequency scale below 0.3 kHz. Even though this is more than two orders of magnitude larger than the first protocol detailed in the main text, it still negligibly contributes to the global uncertainty on the methanol resonance frequency (see Table 1).

## 2 Methanol level and transition notation

In this work, we adopt the notations of [49] for the spectroscopic assignment of methanol lines and transitions. Energy levels are labeled as ($\sigma$,$v$,$t$,$K$,$J$), where $\sigma$ is the torsional symmetry (*A* or *E*), $v$ is the vibrational state, $t$ is the torsional mode ($\nu_{12}$) quantum number, $K$ is the quantum number for the projection of the total orbital angular momentum onto the molecular symmetry axis and $J$ is the total orbital angular momentum quantum number. *K*-doublets of *A* symmetry have +/- superscript on *K* to distinguish the $A^+$ or $A^-$ component of the doublet. For *E* levels, *K* is a signed quantum number with positive (respectively negative) values corresponding to levels often denoted as $E_1$ (respectively $E_2$). Since in this work we only encounter the ground or the first excited state of a few vibrational modes, the quantum number $v$ is expressed by a two-letter label: gr for the ground state, co for the CO-stretching mode ($\nu_8$), sb for the CH$_3$ symmetric bending mode ($\nu_5$) and ro for the CH$_3$ out-of-plane rocking mode ($\nu_{11}$). Simple $\Delta K = 0$ and $\Delta t = 0$ transitions between the ground vibrational state and the first excited state of the vibrational mode $v$ are labeled $P(\sigma,v,t,K,J)$, $Q(\sigma,v,t,K,J)$ or $R(\sigma,v,t,K,J)$ with $\sigma$, $K$ and $J$, the torsional symmetry and rotational quantum numbers of the lower state, and with *P*, *Q* and *R* referring to $\Delta J = -1$, 0 and +1 transitions respectively. $\Delta K = -1$ – respectively +1 – transitions are labeled $^PP(\sigma,v,t,K,J)$, $^PQ(\sigma,v,t,K,J)$ or $^PR(\sigma,v,t,K,J)$ – respectively $^RP(\sigma,v,t,K,J)$, $^RQ(\sigma,v,t,K,J)$ or $^RR(\sigma,v,t,K,J)$. Other transitions are labeled with their upper and lower states ($\sigma$',$v$',$t$',$K$',$J$')←($\sigma$'',$v$'',$t$'',$K$'',$J$'').

## Funding

This work has been supported through EMPIR project 15SIB05 "OFTEN" and the results partly come from the EMPIR project 15SIB03 "OC18". This project has received funding from the EMPIR programme co-financed by the Participating States and from the European Union's Horizon 2020 research and innovation programme. We acknowledge funding support from the Agence Nationale de la Recherche (ANR-15-CE30-0005-01 and Labex First-TF ANR 10 LABX 48 01), Région Ile-de-France (DIM Nano-K and DIM SIRTEQ), CNRS and Université Paris 13. D.B.A. Tran is supported by the Ministry of Education and Training, Vietnam (Program 911).

†These authors contributed equally to this work.
§permanent address: Faculty of Physics, Ho Chi Minh City University of Education, Ho Chi Minh City, Vietnam
#present address: Laboratoire Kastler Brossel, Sorbonne Université, CNRS, ENS-PSL Research University, Collège de France, 4 place Jussieu, 75252 Paris, France
††permanent address: Institute of Laser Physics, Siberian Branch of the Russian Academy of Sciences, Novosibirsk, Russia